\documentclass[prd,aps,showpacs]{revtex4}

\usepackage{graphicx}

\begin{document}

\bibliographystyle{prsty} 

\title{Radiation tails and boundary conditions for black hole
evolutions}

\author{Elspeth W.~Allen}
\author{Elizabeth Buckmiller}
\author{Lior M.~Burko}
\author{Richard H.~Price} 
\affiliation{Department of Physics,
University of Utah, Salt Lake City, Utah 84112}  

\begin{abstract}
\begin{center}
{\bf Abstract}
\end{center}
In numerical computations of Einstein's equations for black hole
spacetimes, it will be necessary to use approximate boundary
conditions at a finite distance from the holes. We point out here that
``tails,'' the inverse power-law decrease of late-time fields, cannot
be expected for such computations.  We present computational
demonstrations and discussions of features of late-time behavior in an
evolution with a boundary condition.
\end{abstract}

\maketitle
\section{Introduction}\label{sec:intro} 
There is at present great interest in the computation of the
gravitational waves from the inspiral and merger of a pair of mutually
orbiting black holes\cite{numrel1,numrel2,numrel3}.  To do such
computations a solution of the initial value equations of general
relativity is chosen on some initial spatial hypersurface, and the
remaining Einstein equations are used to find the spacetime to the
future of that initial surface. In principle, one can compute the
evolved spacetime only in the domain of dependence of that initial
surface, which means that the initial surface must be large, many
times the radius of the initial binary orbit, if the spacetime is to
be evolved for several orbital times. The computational demands for
such a procedure make it unfeasible for the foreseeable future,
although pseudospectral codes can help considerably in extending the
size of the initial hypersurface\cite{scheeletal}.  The alternative to
a large initial hypersurface is a timelike boundary on the
computation, typically at some large radius, at which appropriate
approximate boundary conditions are specified. These boundary
conditions are chosen to represent (approximately) the condition that
no information moves inward through the boundary. One of the problems
that workers in this field have recently turned to is that of
appropriate boundary conditions, especially in connection with the
preservation of gauge constraints\cite{LSU, frittelli,winicour}.

One of the features found in evolutions of perturbations in black
holes spacetimes is the final latest-time behavior of the perturbation
fields, a fall off in time $t$ as $1/t^n$ at a constant distance from
the hole\cite{price, barack}. This was first demonstrated for
Schwarzschild holes in which case $n=2\ell+3$ for a multipole of index
$\ell$ with compact initial support. For Kerr holes such tails also
represent the latest time behavior, though there remains some
controversy about the value of $n$\cite{krivan,poisson,burko_khanna}.
Most work on these tails has been done within linearized perturbation
theory, though some computations with self-gravitating spherically
symmetric scalar fields have also been carried
out\cite{burko_ori,gundlachetal,scheeletal2}.

The purpose of the present
paper is to point out that numerical computations with approximate
boundary conditions cannot be expected to produce the correct
late-time tails.
A rough intuitive reason for this is that the late-time tails are
produced by the scattering of radiation due to the curvature of
spacetime. This scattering takes place far from the hole, and depends
on the asymptotic large-distance nature of spacetime curvature. A
boundary condition on a timelike surface at finite radius means that
the asymptotic nature of the distant spacetime does not enter into the
computation, so that the correct late-time tails cannot develop.

These ideas can be checked accurately in the Schwarzschild
background. In this case linearized perturbations (scalar, electromagnetic,
or gravitational) can be 
analyzed into multipoles
$\psi=\sum\psi_{\ell m}(t,r)Y_{\ell m}(\theta,\phi)$, where $t,r,\theta,\phi$
 are the standard  Schwarzschild coordinates.
 The evolution of
each multipole is described by a simple wave equation\cite{price}
\begin{equation}\label{waveq} 
\partial_t^2\psi(t,r)-\partial_{r*}^2\psi(t,r)+V(r)\psi(t,r)=0\;.
\end{equation}
Here $c=G=1$ and, for simplicity, we have dropped the multipole
indices on $\psi_{\ell m}(t,r)$ and on $V_\ell(r)$. The
Regge-Wheeler\cite{RW} ``tortoise coordinate'' $r^*$ is defined by
$r^*=r+2M\ln(r/2M-1)$, where $M$ is the mass of the Schwarzschild
hole. In the limit $M\rightarrow0$, that is, in flat spacetime,
$r=r^*$ and the potential takes the form
\begin{equation}\label{centripot} 
V^{\rm flat}
= \ell(\ell+1)/r^{*2}\ .
\end{equation}
For this potential, the solution to Eq.~(\ref{waveq}) has a simple 
familiar form in terms of spherical Bessel functions, and has  
no tails. 

\begin{figure}[ht] 
\includegraphics[width=4in]{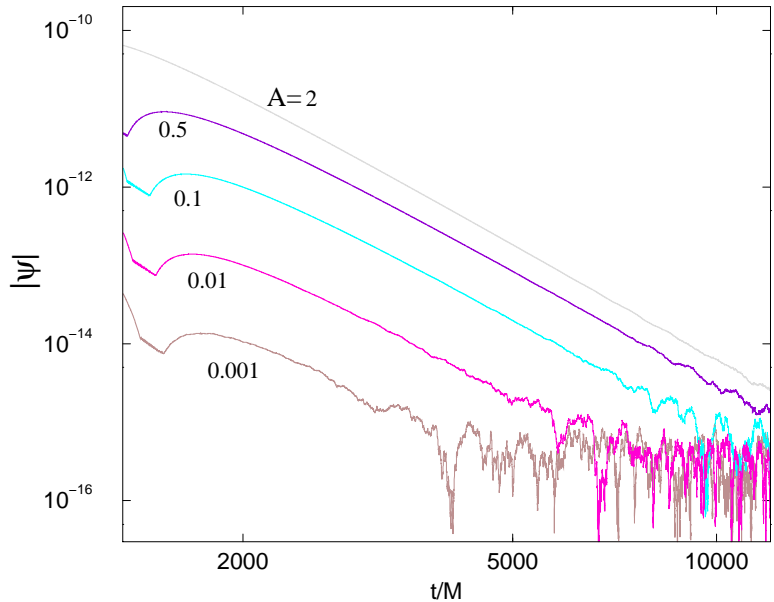}
\caption{
Late-time evolved solutions for different values of $A$,
the coefficient of the scattering term  
in the toy potential described in the text. 
The absolute value of the scalar field $\psi$ is shown as a function
of time $t$ at $r^*=400M$.
Curves are labeled 
with the value of $A$.\label{fig:varA}}
\end{figure}

If $M\neq0$ there are important differences.
For a scalar perturbation, the potential,
\begin{equation}\label{scpot} 
V^{\rm sc}(r)
=\left(1-\frac{2M}{r}\right)\,\left[\frac{\ell(\ell+1)}{r^2}
+\frac{2M}{r^3}\right]\,,
\end{equation}
is typical
of that for all perturbation fields.
For $r^*\gg M$ this potential has the form
\begin{equation}\label{asymptsc} 
V^{\rm sc}
=\frac{\ell(\ell+1)}{r^{*2}}\left[1+4M\;
\frac{\ln (r^*/2M)}{r^{*}}+{\cal O}(M/r^*)
\right]\,.
\end{equation}
It is the extra $\ln r^*/r^*$ term that is in Eq.~(\ref{asymptsc}), but
missing in Eq.~(\ref{centripot}), that produces the $1/t^{2\ell+3}$
tails\cite{chingetal}. To check this we define a toy potential
\begin{equation}
V^{\rm toy}
=\frac{\ell(\ell+1)}{r^{*2}}\left[1+2MA\;
\frac{\ln(r^*/2M)}{r^{*}}
\right]\,,
\end{equation}
with an adjustable parameter, $A$, that allows us to control the size
of the extra term.  Using this potential we have numerically evolved
Eq.~(\ref{waveq}) with $\ell=1$\cite{why1}. The initial data, at time
$t=0$, for this evolution was a time-symmetric Gaussian pulse
$\psi=\exp(-0.1(r^*/2M+4)^2)$. The results, for $\psi$ at $r^*=400M$,
are shown as a function of time $t$ in 
Fig.~\ref{fig:varA}.
The figure shows the straight lines in the log-log plot that indicate
a power-law fall off in time.  For all values of $A$ the slope is -5,
consistent with the $2\ell+3$ rule. (At very late times each of the
power-law tails disappears into round-off noise.)  The results clearly
show the sensitivity of the tail to the size of $A$. For $A=0$, of
course, the toy potential becomes the flat spacetime potential
centrifugal, and there is no tail. (The magnitude of the tail 
increases as  $A$ is made larger, but this increase slows, and
the magnitude of the  tail appears to reach a limit.)


The argument against tails  with finite boundary conditions, then, is
that they depend on the asymptotic form of the potential.  For the
obvious Sommerfeld outgoing boundary condition
\begin{equation}\label{sommerfeld} 
\psi_{,t}+\psi_{,r^*}=0\,, \mbox{ at $r^*=r^*_{\rm bc}$\,,}
\end{equation}
the interior solution cannot ``know'' what the potential is for
$r^*>r^*_{\rm bc}$, and hence the solution cannot develop a tail
except at spacetime points whose domain of dependence lies within the
outer boundary.

This argument, of course, is only suggestive. One can rebut it with
the claim that  a boundary condition could 
be made sufficiently precise so that it encodes the asymptotic
form of the potential. A trivial example is the case $V=0$ and $r=r^*$,
for which the general solution, in terms of an arbitrary outgoing function $f$,
and an arbitrary ingoing function $g$,
is
\begin{equation}\label{monform} 
\psi=f(t-r)+g(t+r)\,.
\end{equation}
In this case 
the
outgoing boundary condition $\psi_{,t}-\psi_{,r}=0$
{\em is} exact; it constrains $g$ to be zero, the same condition
as if the boundary were infinitely far away.
 A less trivial, but less useful case is  the $M=0$,
$\ell=1$
scalar equation . The general solution in this case is 
\begin{equation}\label{dipform} 
\psi={f'(t-r)}+\frac{f(t-r)}{r}\;+\;
{g'(t+r)}-\frac{g(t+r)}{r}\,,
\end{equation}
where $f$ describes the outgoing part and $g$ the ingoing part. 
An exact boundary condition is
\begin{equation}\label{exactdipbc} 
\left(\partial_r+\partial_t\right)^2
\left[r^2\left(\partial_r+\partial_t\right)\psi\right]
=0\,,
\end{equation}
which constrains $g$ (more precisely 
$g^{'''}$) to vanish.

The cases in Eqs.~(\ref{monform}) and (\ref{dipform}) are special. In
these cases the radiative part of the solution (the parts without
$1/r$ factors) and the nonradiative parts are simply related. The
nonradiative part is missing for Eq.~(\ref{monform}), and for
Eq.~(\ref{dipform}) the nonradiative part is a simple time intergral
over the radiative part.  This simplicity is related to Huygens'
principle. Note in particular that in Eq.~(\ref{monform}) or
(\ref{dipform}) if $g=0$ and if $f$ has compact support, then the
solution will be nonzero only for a finite time, and hence cannot have
a power-law tail as a generic feature.
The nonsimple non-Huygens solution for problems in which tails
{\em are} generic will not, therefore, have an exact boundary
condition that can be expressed as a finite number of differentiations
as in Eq.~(\ref{exactdipbc}).

\section{Evolutions with approximate boundary conditions}
We now directly investigate the late-time behavior of solutions with
approximate outgoing boundary condition of Eq.~(\ref{sommerfeld}).
The monopole $\ell=0$ case is studied so that the predicted slowly-falling
 $1/t^3$ tails stay well above the roundoff noise.  The starting
field is  a purely outgoing pulse defined on the ingoing
null ray $t+r^*=0$. On this ray, 
in terms of retarded time $u\equiv t-r^*$,
the form of $\psi$ is specified to
be
$\psi=\left[u(u-8M)/16M^2\right]^8$, 
for $0<u<8M$, and $\psi=0$ for $u<0$ or $u>8M$.
 The initial data at $t=0$ then is approximately an outgoing
pulse (satisfying $\psi_{,r^*}=-\psi_{,t}$) confined between $r^*=-8M$
and $r^*=0$, centered at $r^*=-4M$ with a peak value $\psi=1$ there.
The left boundary for all $\ell=0$ computations is at $r^*=-500M$,
where the sommerfeld condition $\psi_{,r^*}=\psi_{,t}$ is imposed. The
right boundary is placed at different locations $r^*_{\rm bc}$, and
the condition $\psi_{,r^*}=-\psi_{,t}$ is imposed.

\begin{figure}[ht] 
\includegraphics[width=2.5in]{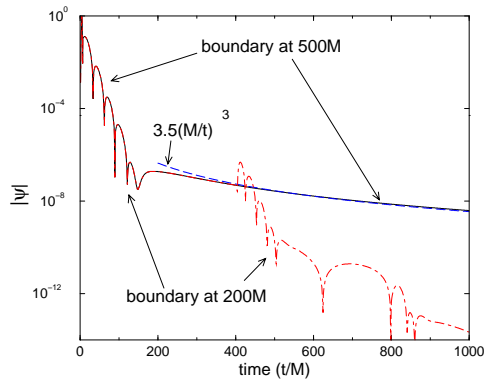} \caption{ The scalar monopole
$\psi$ evolving from an initially outgoing pulse near $r^*=0$. The
field $\psi$ is plotted as a function of time at $r^*=0$ for two
different values of $r_{\rm bc}^*$, at $500M$ and $200M$.  The two
curves are identical up to time $t=400M$.  A curve proportional to
$1/t^3$ is included for comparison.  \label{fig:farbc}}
\end{figure}

Figure \ref{fig:farbc} shows the time profile of $\psi$ developing
from the initial pulse. The value of $\psi$ at the fixed position
$r^*=0$ is shown as a function of time for outgoing boundary locations
$r^*_{\rm bc}=500M$ (solid curve) and at $r^*_{\rm bc}=200M$
(dot-dashed curve). Since the results are shown only up to $t=1000M$,
the field at $r^*=0$, shown by the solid curve, has not yet been
influenced by the interaction of the boundaries at $\pm500M$ with the
evolution of the initial pulse from $r^*=0$. The solid curve then
shows the ``boundaryless'' behavior of the field evolving completely
within the domain of dependence of the initial data.  That field goes
through quasinormal ringing up to around $t=150M$, then around
$t=300M$ becomes a power-law tail. Since the plot is a semilog plot,
not a log-log plot, the tail is not a straight line, and a ${\rm
constant}/t^3$ curve is provided to illustrate the tail nature of the
solid curve.
The dot-dashed curve for the $r^*_{\rm bc}=200M$ outer boundary is
identical to the solid curve up to $t=400M$. In particular, from
$t\approx300M$ to $t=400M$ the field starts to take the form of a
power-law tail, but at $t=400M$ waves reflected from the $r^*_{\rm
bc}=200M$ boundary arrive at $r^*=0$ and a new oscillatory behavior
ensues due to the influence of the boundary.

\begin{figure}[ht] 
\includegraphics[width=2.5in]{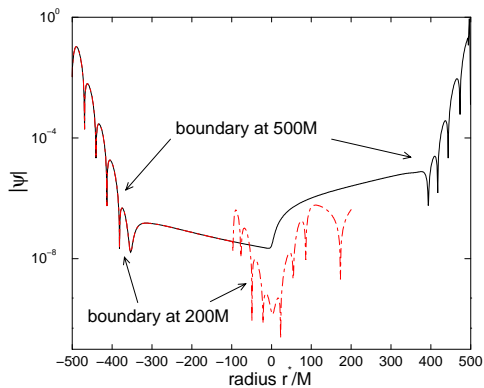}
\caption{Spatial profile of 
the scalar monopole $\psi$ evolving from an initially 
outgoing pulse. Here $\psi$ is plotted at $t=500M$
for values of $r_{\rm bc}^*$ at $500M$ and $200M$.
The curves are identical for $r^*\leq-100M$ where the 
field has not yet been influenced by reflections of 
the initial pulse off the $200M$ boundary.
\label{fig:myspatial}}
\end{figure}

A complementary viewpoint on the boundary influence is given in
Fig.~\ref{fig:myspatial}, in which spatial profiles of evolved 
fields are shown at the moment of time $t=500M$. As in
Fig.~\ref{fig:farbc}, the solid curve corresponds to an outer boundary
at $r^*=500M$, and the dot-dashed curve to an outer boundary at
$r^*=200M$.  The solid curve indicates that at $t=500M$ the
unit-height pulse has reached the outer boundary. In the absence of
scattering this would be the only feature of the plot, but due to
scattering there are quasinormal bumps created between
$r^*\approx370M$ and $500M$. Waves scattered inward also create
quasinormal bumps between $r^*\approx-300M$ and $-500M$. From
$r^*\approx-300M$ to $\approx 370M$ the solid curve shows the spatial
profile of the tail behavior.
The dot-dashed curve shows that reflections of the inital pulse off
the $r^*=200M$ boundary have reached $r^*=-100M$ and have
``contaminated'' the spatial profile from $ r^*=-100M$ to $200M$.  The
spatial profile from $ r^*=-100M$ to $200M$ shows spatial oscillations
similar to the temporal oscillations that appear in
Fig.~\ref{fig:farbc}.

\section{Boundary-induced quasinormal ringing}

\begin{figure}[ht] 
\includegraphics[width=3.5in]{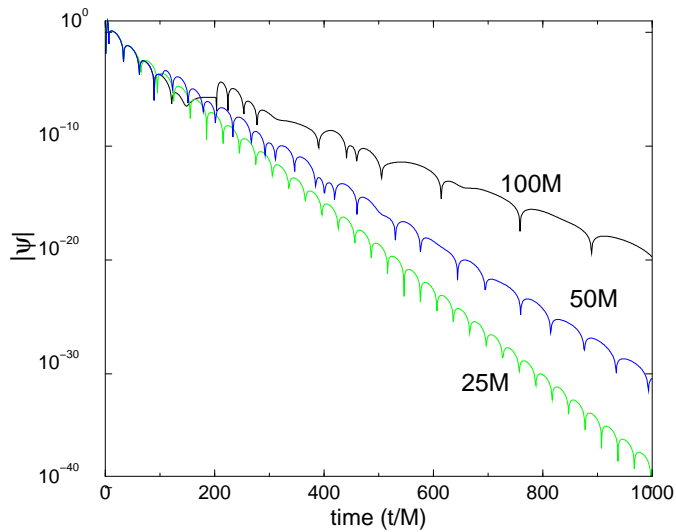} \caption{ Evolution of the $\psi$
monopole for Sommerfeld outgoing conditions on close boundaries. The
time profiles are for an initial outgoing pulse near $r^*=0$. Results
are shown for outer boundaries at locations $r*=100M$, $50M$, and
$25M$.  \label{fig:bumps}}
\end{figure}

The nature of the boundary-induced oscillations is made clearer in
Fig.~\ref{fig:bumps}, which shows the time profiles for outer
boundaries at $100M$, $50M$, $25M$.  (For larger values of $r^*_{\rm
bc}$ similar oscillations develop at later times.)  The late-time
results for the $50M$ and $25M$ profiles indicates a constant-period
oscillation, and the straight-line envelope of the oscillations in the
semilog plot indicates an exponential damping. The natural
interpretation is that these damped oscillations are a new form of
quasinormal ringing.  The ``old'' form is the familiar quasinormal
ringing of a black hole\cite{vishnature,livingrev,nollert}, like that
in Fig.~\ref{fig:farbc} for $t/M$ less than $\sim150$.
This is a real physical phenomenon associated with the black hole
spactime.   In Fig.~\ref{fig:bumps} we see oscillations that are
not physical in this sense, but 
are  numerical artifacts, the strongly damped oscillations of a
leaky cavity created by the outer boundary and the 
curvature potential $V$ near its peak at $r^*=0$. Some evidence for
this is the dependence of the period of the oscillations on $r^*_{\rm
bc}$. The longer cavity of the $50M$-boundary case creates longer
oscillation periods than the cavity of the $25M$-boundary. The
less-clearly defined oscillations of the $100M$ have yet longer
periods.

\begin{figure}[ht] 
\includegraphics[width=3in]{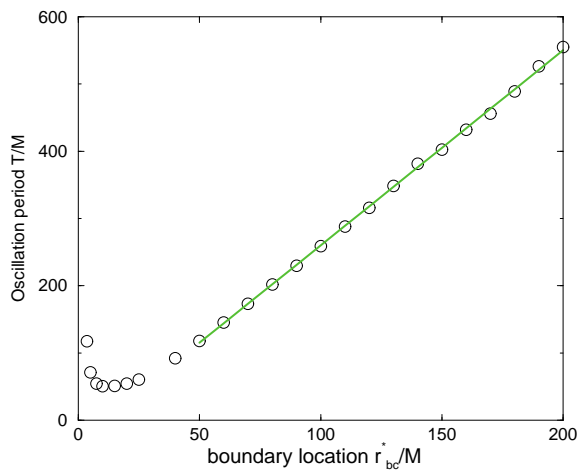} \caption{ 
The period $T$ for a full oscillation (two ``bumps'')
as a function of the location $r^*_{\rm bc}$ of the 
outer boundary. The straight line $T/M=-30+2.9(r^*_{\rm bc}/M)$
is the least-squares fit to the results 
for $r^*_{\rm bc}
=100M$ and larger. \label{fig:pervsrbc}}
\end{figure}

The leaky-cavity interpretation is supported by the plot in
Fig.~\ref{fig:pervsrbc} showing the period $T$ for a full oscillation
(the width of a pair of ``bumps'' in Fig.~\ref{fig:bumps}) of the late
time features, as a function of $r^*_{\rm bc}$, the location of the
outer boundary. The uncertainty in the period is around 2\%, both due
to truncation error in the evolution of $\psi$, and due to the
extraction of $T$ from the late time results.
For $r^*_{\rm bc}$ larger than $\sim100M$ the period
quite accurately follows the linear relation $T/M=-30+2.9(r^*_{\rm
bc}/M)$, confirming the view that these late time oscillations are
resonances of a cavity created by the outer boundary. For $r^*_{\rm
bc}$ less than $\sim100M$ it is not surprising that the details near
the peak of the curvature potential $V$ would complicate the
relationship; for large values of $r^*_{\rm bc}$ the details of the
curvature potential become unimportant.  What {\em is}
at first surprising in the linear relationship is the coefficient
$2.9$. This means that the number of half-wavelengths inside the
``cavity'' is not an integer; one would naively expect the period $T$
to be equal to the cavity length, or an integer multiple of half the
cavity length. But our naive expectations are based on the simple
boundary condition that the field or its normal derivative vanish. In
our artificial cavity the outgoing boundary conditions does
neither. For a constant-frequency oscillation the outgoing boundary
condition, in effect, imposes a relationship between the field and its
normal derivative. Computations with simple toy models have confirmed
that with such boundary conditions a resonance of the cavity will not
contain an integer number of half wavelengths.

\begin{figure}[ht] 
\includegraphics[width=3in]{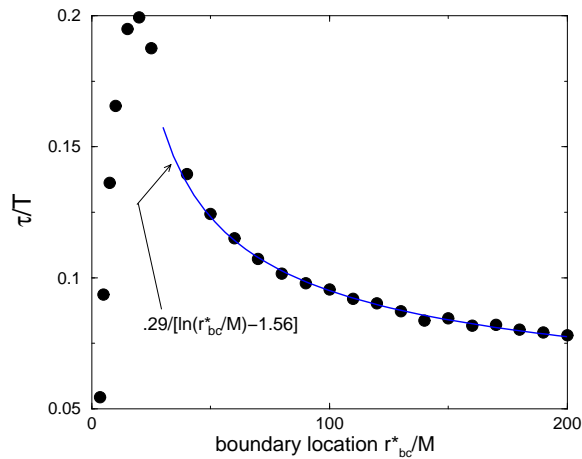} \caption{ The time constant
$\tau$ for the envelope of the decaying oscillations, divided by $T$,
the period of the oscillations, as a function of the location
$r^*_{\rm bc}$ of the outer boundary.  The thick solid curve shows a
heuristic model for $\tau$. \label{fig:taubyT}}
\end{figure}

The leaky-cavity viewpoint on the outer boundary suggests a way to
quantify the effectiveness of the outer boundary condition. The more
effective the outer boundary condition is, the more quickly the cavity
modes should die out. If we characterize the effective reflectivity of
the outer boundary as $R$, and let $N$ represent the number of
reflections from the outer boundary, then the amplitude should fall
off as $R^N$. We can take $N$ to be the time $t$ divided by some
measure of the time for a reflection. For generality we will take this
to be $\kappa_1T$, were $\kappa_1$ is some constant, and $T$ is the
period for an oscillation. The cavity oscillations should then die off
as $R^{t/(\kappa_1 T)}$, or as $\exp(-t/\tau)$, where the damping time
$\tau$ is $-T\kappa_1/\ln(R)$. The outgoing boundary condition is
expected to improve as $r^*_{\rm bc}$ increases\cite{reflectivity}, so
it is interesting to make the {\em ansatz} $R=\kappa_2 M/r^*_{\rm
bc}$. This model predicts $\tau/T=\kappa_1/\left[\log(r^*_{\rm
bc}/M)-\log(\kappa_2)\right]$.  Figure \ref{fig:taubyT} shows $\tau/T$
for a range of outer boundary locations $r^*_{\rm bc}$; the
uncertainty here, as in Fig.~\ref{fig:pervsrbc}, is no worse than 2\%
at any $r^*_{\rm bc}$. The large-$r^*_{\rm bc}$ results in
Fig.~\ref{fig:taubyT} do show a gradual decrease of $\tau/T$ with
increasing $r^*_{\rm bc}$, as expected.  Our heuristic model for
$\tau$ is fit to those results by eye, and is plotted in
Fig.~\ref{fig:taubyT}. That fit corresponds to reflectivity
$~4.8M/r^*_{\rm bc}$, or $\sim5$\% for a boundary at $100M$.  The plot
of the heuristic model gives an appearance of reasonable agreement
except for small values of $r^*_{\rm bc}$, but with two adjustable
parameters in the fit, this agreement can only be said to be weakly
suggestive.

\section{Summary and discussion}

We have shown that finite-radius boundary conditions prevent the
formation of power-law tails of perturbations in black hole
spacetimes. In place of a tail, the latest-time feature of a
computation will be a new form of quasinormal oscillation. Unlike
black hole oscillations, these oscillations are not physical
phenomena; they are numerical artifacts introduced by the imperfect
outgoing condition at the outer boundary.

In numerical relativity, these oscillation will probably not be a
serious practical difficulty.  The goal of present numerical
relativity work is a better understanding of strong field nonlinear
dynamics. Numerical codes in the forseeable future will not be able to
run long enough for the boundary-induced oscillations to appear, nor
are they likely to be accurate enough to deal with such a weak field
phenomenon.

These limitations of running time and accuracy do not apply when the
Lazarus\cite{lazarus} method is used in a problem involving the
formation of a final black hole. That method uses the solution
computed by a fully nonlinear numerical evolution code as initial data
for further evolution by black hole perturbation theory.  The
nonlinear numerical evolution would, of necessity, use timelike
boundaries, but the numerical cavity oscillations would not develop in
the limited time for which the evolutions run.  In principle, the
subsequent Lazarus evolution could be used without boundary
conditions, i.e., with evolution only within the domain of dependence
of the initial data inherited from the fully nonlinear numerical
code. Such boundaryless Lazarus evolutions would exhibit power-law
tails, but the tails would be strongly affected by boundary effects
contained in the initial data inherited from the nonlinear
evolution. In practice, Lazarus evolutions are not
boundaryless. Rather, to reduce memory requirements, timelike boundary
conditions are used in the Lazarus perturbation evolutions.  These
evolutions should contain boundary-induced artifacts of the type we
have discussed above. But these artifacts would be miniscule, and of
no concern for most applications of black hole evolution, with or
without Lazarus.

\section{Acknowledgment} We gratefully acknowledge the support of the
National Science Foundation to the University of Utah, under grants
PHY9734871 and PHY0244605. We thank Gioel Calabrese, Manuel Tiglio,
Luis Lehner and Jorge Pullin  for discussions and 
suggestions about boundary effects. We thank Carlos Lousto for 
discussions of the boundary treatment in the Lazarus method.





\begin{thebibliography}{99}
\bibitem{numrel1} S.~Brandt {\em et al.},  
Phys.~Rev.~Lett. {\bf85}, 5496 (2000).
\bibitem{numrel2} 
M.~Alcubierre {\em et al.}, Phys.~Rev.~Lett. {\bf87}, 271103 (2001).
\bibitem{numrel3}L. Lehner, Class.~Quantum Grav. {\bf 18}, R25 (2001). 





\bibitem{scheeletal} 
M.~Scheel {\em et al.}, Phys.~Rev.~D {\bf66}, 124005 (2002).
 

\bibitem{LSU} 
G.~Calabrese, L.~Lehner, and M.~Tiglio, Phys.~Rev.~D {\bf65}, 
104031 (2002).

\bibitem{frittelli} 
S.~Frittelli and R.~Gomez, Class.~Quantum Grav. {\bf20}, 
2379 (2003).

\bibitem{winicour} B.~Szilagyi and J.~Winicour, 
 Phys.~Rev.~D{\bf68},  041501 (2003).

\bibitem{price} R.~H.~Price, Phys. Rev. D{\bf 5}, 2419 (1972).

\bibitem{barack} 
L.~Barack, Phys.~Rev.~D{\bf 59}, 044016 (1999); Phys.~Rev.~D{\bf 59},
044017 (1999).
  
\bibitem{krivan} W. Krivan, Phys. Rev.~ D {\bf60}, 101501 (1999).

\bibitem{poisson} 
E. Poisson, Phys. Rev.~ D {\bf66}, 044008 (2002).

\bibitem{burko_khanna} 
L.~M.~Burko and G.~Khanna, 
Phys.~Rev.~D{\bf67} 081502, (2003).

  \bibitem{gundlachetal} 
C.~Gundlach, R.~H.~Price and J.~Pullin,
Phys.~Rev.~D{\bf 49}, 890 (1994).


\bibitem{burko_ori}
L.~M.~Burko and A.~Ori, 
 Phys.~Rev.~D{\bf56},  7820 (1997). 
 
  \bibitem{scheeletal2} 
M.~A.~Scheel {\em et al.} preprint gr-qc/0305027.


\bibitem{RW}T.~Regge and J.~A.~Wheeler, Phys.~Rev. {\bf108}, 1063 (1957).

\bibitem{chingetal} E.~S.~C.~Ching {\rm et al.},
Phys.~Rev.~Lett. {\bf74}, 2414 (1995).


\bibitem{why1} Our subsequent computations will use $\ell=0$, but
here we use $\ell=1$ to make it clear that there can be a nonzero
potential (the $A=0 $, $\ell=1$ case) without tails.




\bibitem{vishnature}C.V. Vishveshwara,  {\it Nature},
  {\bf227}, 936 (1970).


\bibitem{livingrev} K. D. Kokkotas and B. G. Schmidt, Living Rev.~Relativ. {\bf2}, 2 (1999).

\bibitem{nollert} H.-P. Nollert, Class.~Quantum Grav. {\bf16}, R159 (1999). 

\bibitem{reflectivity} The usual argument is that 
the error in the outgoing boundary condition is of order ${\rm
wavelength}/r^*_{bc}$. The wavelength of our ``cavity oscillations''
is proportional to $r^*_{bc}$. It would be consistent, therefore, to
suggest that $\tau/T$ should be independent of $r^*_{bc}$, and our
numerical results are more-or-less compatible with this.


\bibitem{lazarus}J. Baker, M. Campanelli, and C.~Lousto,
Phys.~Rev.~D.{\bf65}, 044001 (2002). 

\end{thebibliography}
\end{document}